\documentclass[9pt,twocolumn,twoside]{osajnl}

\journal{ol} 
\usepackage{siunitx}

\setboolean{shortarticle}{true}

\title{Topological valley plasmon transport in bilayer graphene metasurfaces for sensing applications}

\author[1]{Yupei Wang}
\author[1]{Jian Wei You}
\author[1]{Zhihao Lan}
\author[1,*]{Nicolae C. Panoiu}
\affil[1]{Department of Electronic and Electrical Engineering, University College London,
    Torrington Place, London WC1E 7JE, United Kingdom}
\affil[*]{Corresponding author: n.panoiu@ucl.ac.uk}

\begin{abstract}
Topologically protected plasmonic modes located inside topological bandgaps are attracting
increasing attention, chiefly due to their robustness against disorder-induced backscattering.
Here, we introduce a bilayer graphene metasurface that possesses plasmonic topological valley
interface modes when the mirror symmetry of the metasurface is broken by horizontally shifting in
opposite directions the lattice of holes of the top layer of the two freestanding graphene layers.
In this configuration, light propagation along the domain-wall interface of the bilayer graphene
metasurface shows unidirectional features. Moreover, we have designed a molecular sensor based on
the topological properties of this metasurface using the fact that the Fermi energy of graphene
varies upon chemical doping, namely molecular adsorption in our case. This effect induces strong
variation of the transmission of the topological guided modes, which can be employed as the
underlying working principle of gas sensing devices. Our work opens up new ways of developing
robust integrated plasmonic devices for molecular sensing.
\end{abstract}

\setboolean{displaycopyright}{true}

\begin{document}

\maketitle

Research in topological photonics, inspired by the theory of quantum Hall effect in solid-state
physics, has led to the discovery of novel and unique phenomena, such as unidirectional,
defect-immune, and scattering-free propagation of light
\cite{lu2014topological,khanikaev2017two,ozawa2019topological,niu2007prl,shvets2016njp,zhang2017ol},
which have the potential to contribute to the development of robust on-chip ultracompact
nanophotonic devices. Topological photonic modes could be achieved by gapping out
symmetry-protected Dirac cones, for example, through time-reversal symmetry breaking induced by
magneto-optical effects under an external static magnetic field, or spatial-inversion symmetry
breaking induced by spatially asymmetric perturbations
\cite{lu2014topological,khanikaev2017two,ozawa2019topological}. Currently, a variety of
experimental platforms are available for topological photonics, including metamaterials, photonic
crystals, and evanescently coupled waveguides and optical-ring resonators
\cite{ozawa2019topological}.

Valley degree of freedom, which is associated to the conduction-band minima (or valence-band
maxima) in graphene-like two-dimensional (2D) materials \cite{niu2007prl}, has recently been
introduced to photonics \cite{shvets2016njp}, too. These materials exhibit nontrivial Berry
curvature distribution in the momentum space around each valley, which gives rise to a
valley-dependent topological index associated to the integral of Berry curvature around a valley
\cite{niu2007prl}. Furthermore, a domain-wall interface separating two topologically distinct
valley photonic crystals can support valley-momentum locked modes localized at the interface,
similar to the quantum-valley Hall effect \cite{shvets2016njp}. Until now, valley-Hall photonic
modes have been mostly studied in bulk materials, such as photonic crystals
\cite{lu2014topological,khanikaev2017two,ozawa2019topological}, being less explored in 2D photonic
platforms, including graphene \cite{you2020}. This 2D material is becoming a promising platform to
achieve passive and active topologically protected plasmonic modes \cite{jin2017infrared,ylp20sa},
due to its high carrier mobility and long relaxation time
\cite{you2018nonlinear,gonccalves2016introduction}. Equally important, recent advances in
nanofabrication techniques make it possible to achieve graphene based plasmonic nanostructures with
particularly complex geometrical configurations\cite{singh2011graphene,zhang2013review}.

In this paper, we design a bilayer graphene metasurface to realize, to the best of our knowledge
for the first time, valley topological plasmonic modes by utilizing a novel mechanism of mirror
symmetry breaking between the top and bottom freestanding layers of a graphene metasurface, by
horizontally shifting in opposite directions the lattice of holes of the top layer. As such, the
symmetry-protected Dirac cones are gapped out and, consequently, a topological nontrivial frequency
gap emerges. Furthermore, topologically guided valley modes are observed along a domain-wall
interface with respect to which the composite metasurface is mirror symmetric. Our full-wave
numerical simulations, based on solving the full set of 3D Maxwell equations via the finite element
method, verify that the light propagation along the domain-wall interface shows indeed
unidirectional feature. Employing this unique feature of unidirectional propagation and the tunable
optical response of graphene, a molecular sensor based on this newly proposed topological
metasurface is designed and its sensitivity and functionality are quantitatively characterized.
\begin{figure}[t!]
    \centering
    \includegraphics[width=0.8\linewidth]{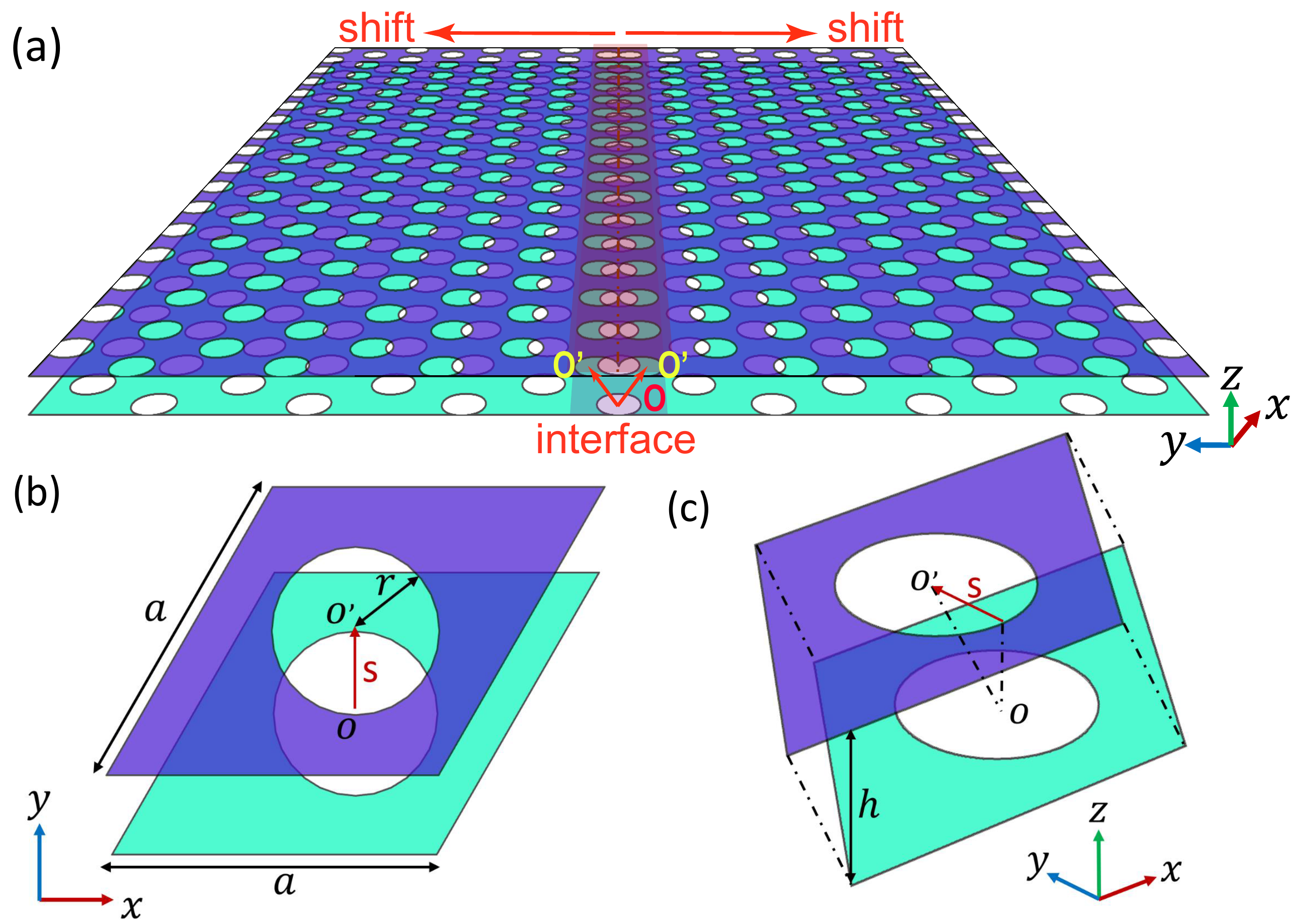}
    \caption{\label{fig:fig1}Schematic of the bilayer graphene metasurface. (a) The metasurface
        contains a domain-wall interface oriented along the $x$-axis, which is constructed by shifting the
        hole lattices of the two halves of the top graphene layer (purple) w.r.t. the bottom layer (green)
        along the positive and negative directions of the $y$-axis. (b) Top view of the unit cell with
        lattice constant, $a$, and horizontal shift, $s$. Hole centers $O$ and $O^{\prime}$ correspond to
        the unit cells of the bottom and top layer, respectively. (c) Bird's eye view of the unit cell with
        a separation distance, $h$, between the two layers.}
    \noindent\rule[0.25\baselineskip]{\linewidth}{0.5pt}
    \vspace{-9mm}
\end{figure}

The schematic of the proposed topological bilayer graphene metasurface is shown in
Fig.~\ref{fig:fig1}. It consists of two freestanding, optically coupled graphene plasmonic crystals
with the same unit cell. Note that the conclusions of this study remain qualitatively valid if one
assumes that the two graphene layers are separated by a certain dielectric material instead of air,
the only changes being of quantitative nature. Moreover, the hole lattice of the left- and
right-hand side domains of the top graphene layer are horizontally shifted, in opposite directions
and normally onto an interface lying along the $x$-axis, by a certain distance, $s$. Each domain
consists of a hexagonal graphene plasmonic crystal with a hole in the unit cell. The top and
bird's-eye views of the unit cell are given in Figs.~\ref{fig:fig1}(b) and~\ref{fig:fig1}(c),
respectively. In this work, we fix the lattice constant $a=\SI{400}{\nano\meter}$, the radius of
holes $r=\SI{100}{\nano\meter}$, and the distance between the top and bottom graphene layers
$h=\SI{90}{\nano\meter}$.

The optical properties of graphene are described by its electric permittivity, $\epsilon_g$, which
is given by Kubo's formula \cite{gonccalves2016introduction}:
\begin{align}
    \epsilon_g(\omega)=&1-\frac{e^2}{4\epsilon_0\pi\hbar\omega h_g}\ln\left(
    {\frac{\xi-i\bar{\omega}}{\xi+i\bar{\omega}}}
    \right)\nonumber\\
    &+\frac{ie^2k_BT\tau}{\epsilon_0\pi\hbar^2\omega\bar{\omega}h_g}\left[
    \frac{\mu_c}{k_BT}+2\ln\left({e^{-\frac{\mu_c}{k_BT}}+1}\right)
    \right]
\end{align}
where $\omega$ is the frequency, $T$ is the temperature, $\mu_c$ is the chemical potential,
$h_g=\SI{0.5}{\nano\meter}$ is the graphene thickness, $\bar{\omega}=1-i\omega\tau$, and
$\xi=2\tau|\mu_c|/\hbar$, with $\tau$ being the relaxation time. Note that dispersive and
dissipative effects are incorporated in our simulations via the frequency-dependent complex surface
conductivity of graphene, defined as $\sigma_s=-i\epsilon_0\omega h_g(\epsilon_g-1)$. In our
analysis, $T=\SI{300}{\kelvin}$ and $\tau=\SI{50}{\pico\second}$, and we set
$\mu_c=\SI{0.2}{\electronvolt}$ unless otherwise stated.

\begin{figure}[t!]
    \centering
    \includegraphics[width=\linewidth]{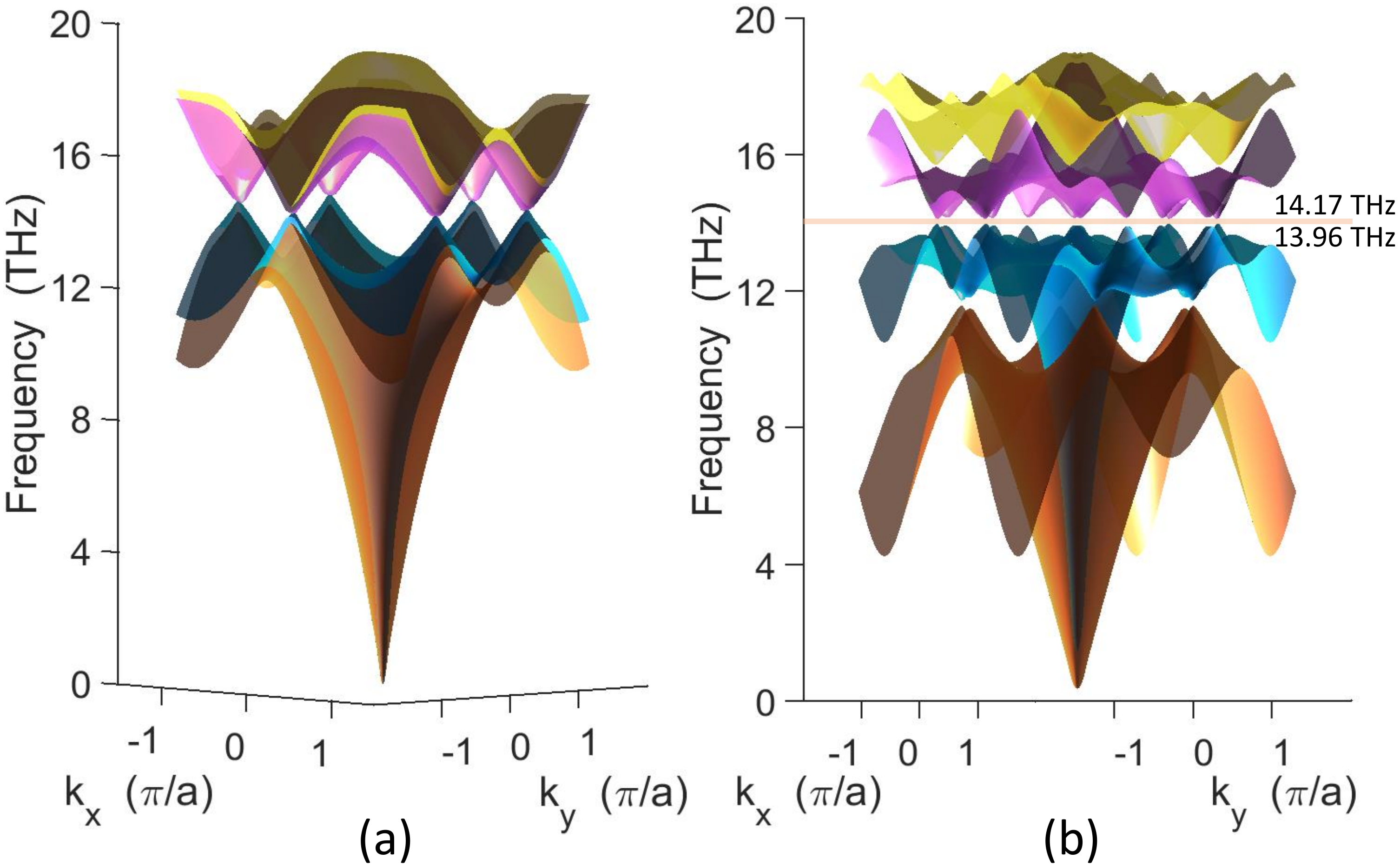}
    \caption{\label{fig:fig2}(a) Band diagram of a bilayer graphene metasurface with $s=0$, in which
        the coupling between the top and bottom graphene nanohole crystals is very weak. (b) Band diagram
        of a composite bilayer graphene metasurface in which the coupling between the top and bottom
        graphene plasmonic crystals is relatively strong, namely $h=\SI{90}{\nano\meter}$ and
        $s=\SI{100}{\nano\meter}$, as depicted in Fig.~\ref{fig:fig1}(c). Since the mirror symmetry of the
        composite graphene metasurface is broken in this case, a nontrivial bandgap corresponding to the
        beige region emerges.}
    \noindent\rule[0.25\baselineskip]{\linewidth}{0.5pt}
    \vspace{-9mm}
\end{figure}

Unlike the case of the mirror-symmetric bilayer graphene metasurface ($s=0$), in the case of a
metasurface with $s\neq0$, the frequency maxima and minima are not necessarily located at the
high-symmetry points of the first Brillouin zone (FBZ) \cite{kittel1996introduction}. Thus, in
order to properly identify the frequency band gap, the plasmonic bands of this bilayer graphene
metasurface have been evaluated in the entire FBZ and the results are given in Fig.~\ref{fig:fig2}
(we have used the Wave Optics Module of COMSOL Multiphysics 5.4). When the distance $h$ between the
top and bottom graphene layers in Fig.~\ref{fig:fig1}(c) is large, the optical near-field coupling
between the two layers can be neglected. As a consequence, each graphene layer, which is a
plasmonic crystal, possesses decoupled Dirac cones protected by $D_{6h}$ point symmetry group
\cite{xie2018photonics}. This is indeed verified by the bands presented in Fig.~\ref{fig:fig2}(a),
where the Dirac cones located at \SI{14}{\tera\hertz} of each graphene plasmonic crystal perfectly
overlap. In order to enhance the optical coupling between the top and bottom graphene layers, the
distance $h$ is reduced to $h=\SI{90}{\nano\meter}$ and, to break the mirror symmetry between the
top and bottom graphene layers, a shift of $s=\SI{100}{\nano\meter}$ is introduced as explained
above. As a result, the $D_{6h}$-symmetry-protected Dirac cones are gapped out, and a frequency
band gap emerges. Specifically, the band diagram of the bilayer graphene metasurface exhibits a
\SI{0.21}{\tera\hertz} topological gap from \SIrange{13.96}{14.17}{\tera\hertz}, as depicted in
Fig.~\ref{fig:fig2}(b). Note that all the results reported here remain qualitatively the same for
smaller $s$ but the frequency bandgap would be narrower. Thus, if $s$ is reduced to
$\SI{70}{\nano\meter}$, the bandgap would decrease by $\sim$\SI{10}{\percent}.

Since the bilayer graphene metasurface has hexagonal symmetry, it possesses six Dirac cones (see
Fig.~\ref{fig:fig2}) with two non-equivalent valleys at $K$ and $K^{\prime}$ symmetry points. The
integral of the Berry curvature around each valley defines the valley Chern number of $C_{K,K'}=\pm
1/2$ \cite{shvets2016njp}. Moreover, the two valleys at $K$ and $K^{\prime}$ are related to each
other \textit{via} rotations of the metasurface by $\pi/3$, $\pi$, and $5\pi/3$. Therefore, in
order to construct a domain-wall interface that can possess topological interface modes, one can
place together two bilayer graphene metasurfaces with $s\neq0$ in a mirror-symmetric manner, i.e.
rotated by $\pi$ w.r.t. each other, as per Fig.~\ref{fig:fig1}(a). Consequently, the difference of
the valley Chern number across the domain-wall interface at each valley is $+1$ or $-1$. In this
way, we can obtain a pair of valley-momentum locked interface states, where the interface state at
one valley has a positive velocity whereas the other has a negative one.

\begin{figure}[t!]
    \centering
    \includegraphics[width=\linewidth]{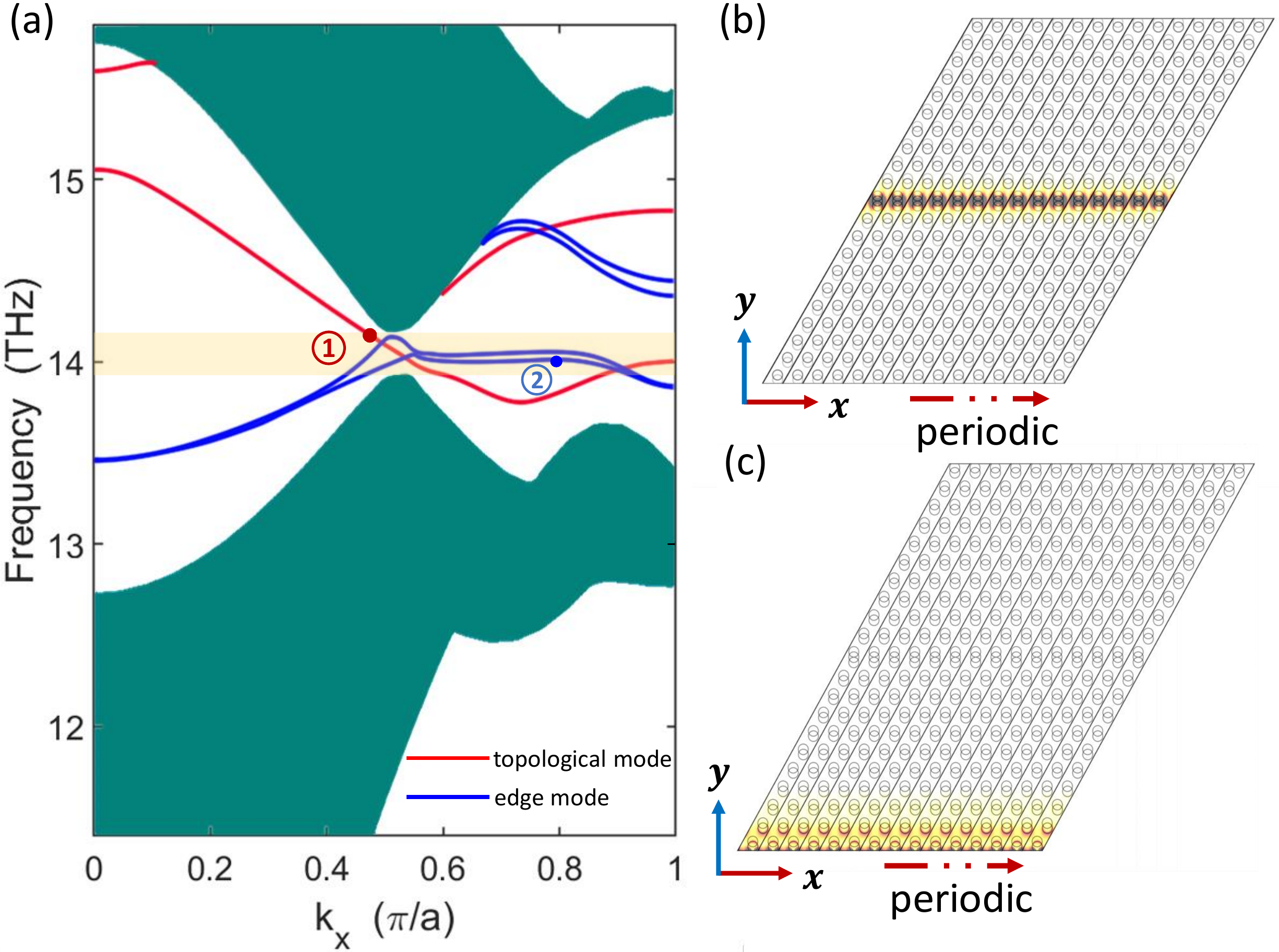}
    \caption{\label{fig:fig3} (a) Projected band diagram (green region), topological interface modes
        (red lines), and non-topological edge modes (blue lines), determined for a finite bilayer graphene
        metasurface with width of 20 unit cells and $s=\SI{100}{\nano\meter}$. (b), (c) Field distributions
        of a topological interface mode and edge mode, marked by \textcircled{1} and \textcircled{2} in
        Fig. 3(a), respectively.}
    \noindent\rule[0.25\baselineskip]{\linewidth}{0.5pt}
    \vspace{-9mm}
\end{figure}

The projected band diagram of a finite bilayer graphene metasurface consisting of 20 unit cells
along the $y$-axis and periodic along the $x$-axis is computed, and the results are presented in
Fig.~\ref{fig:fig3}(a). In this figure, the green regions represent the bulk states and the
topological interface modes are marked by red lines. Note that since the bilayer graphene
metasurface has a finite number of unit cells along the $y$-axis, there are additional edge modes
in Fig.~\ref{fig:fig3}(a). More specifically, the blue lines in this figure indicate edge modes
confined at the metasurface boundaries rather than the domain-wall interface. In particular, we
find that the edge modes generally appear in pairs, at the top and bottom boundaries, and their
influence on the domain-wall interface topological modes can be neglected when the number of unit
cells of each domain along $y$-axis is larger than about $7$. However, additional functionality can
be achieved in a photonic system in which these modes become optically coupled
\cite{tsofe2007pre,ha2008josa}.

In order to gain deeper physical insights into the properties of these interface and edge modes,
their corresponding field distribution are further investigated. To be more specific, the field
distribution of the interface mode \textcircled{1} in Fig.~\ref{fig:fig3}(a), given in
Fig.~\ref{fig:fig3}(b), is highly confined at the domain-wall interface, whereas the field
distribution of the edge mode \textcircled{2} in Fig.~\ref{fig:fig3}(a), presented in
Fig.~\ref{fig:fig3}(c), is confined at the boundary of the finite metasurface.

\begin{figure}[t!]
    \centering
    \includegraphics[width=0.8\linewidth]{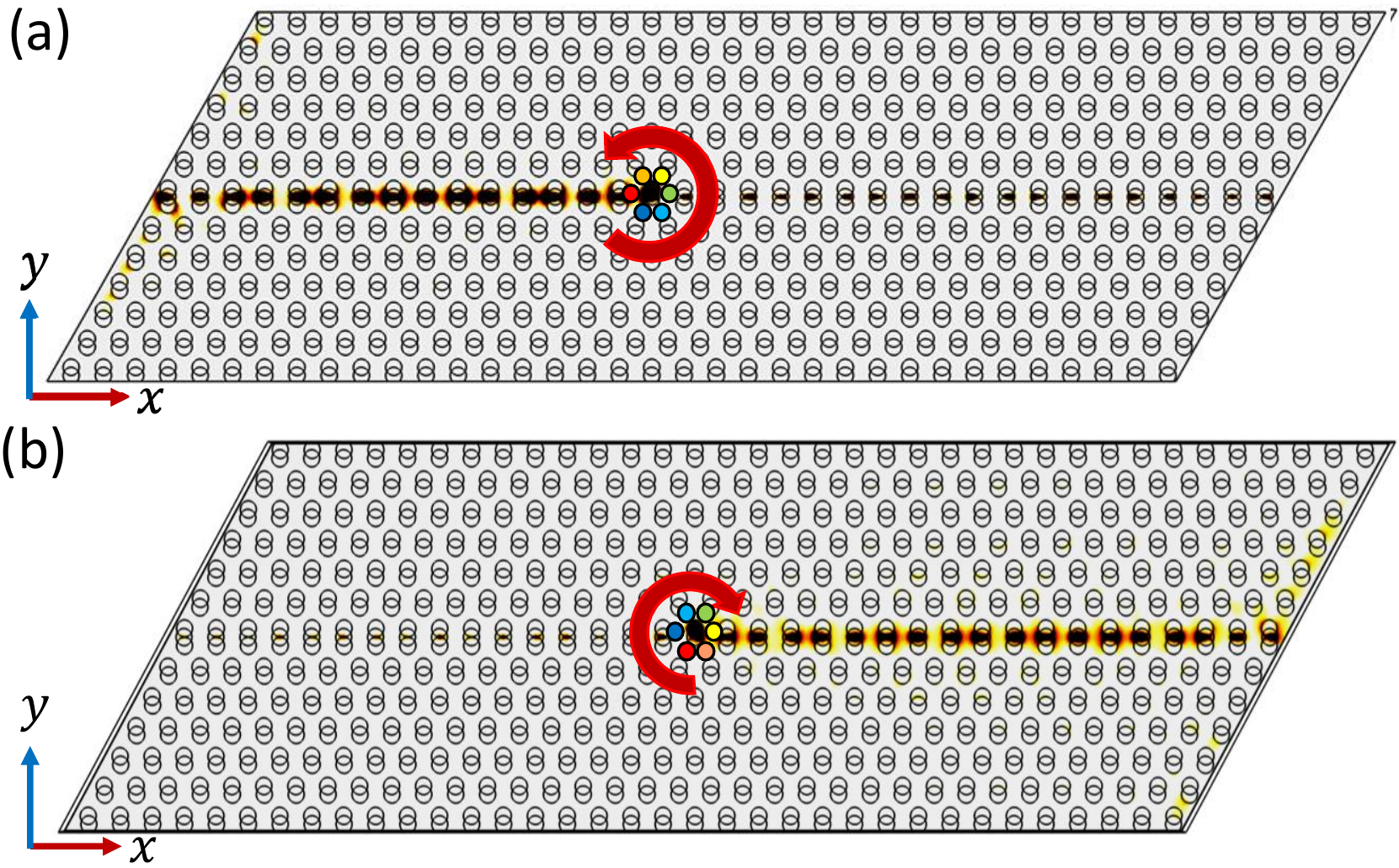}
    \caption{\label{fig:fig4}(a) Unidirectional propagation along the negative direction of the
        $x$-axis, when the finite metasurface is excited by a right-circularly polarized source. (b) The
        same as in (a), but for a left-circularly polarized source. In this case, the topological
        interfacial mode propagates along the positive direction of the $x$-axis.}
    \noindent\rule[0.25\baselineskip]{\linewidth}{0.5pt}
    \vspace{-10mm}
\end{figure}

Importantly, we also demonstrated the unidirectional character of light propagation along the
domain-wall interface. To illustrate this, we simulated the composite bilayer graphene metasurface
with absorbing boundary conditions, but left a certain air space along the $z$-axis to allow for
the radiation losses. A monochromatic light source with frequency of \SI{14.16}{\tera\hertz} was
used to excite the proposed composite bilayer graphene metasurface. In order to study the
chirality-momentum locking property, which arises from the valley-Hall effects induced by the
intrinsic chirality associated to each valley \cite{niu2007prl}, the excitation source was
constructed by placing at the corners of a small hexagon six electric dipoles, marked by circles in
Fig.~\ref{fig:fig4}. In our simulations, the phase difference between neighboring dipoles is set to
$\pm \pi/3$, so as to implement right-circularly polarized (RCP) and left-circularly polarized
(LCP) sources, respectively. More specifically, as illustrated in Fig.~\ref{fig:fig4}(a), a RCP
light source is placed at the center of the composite bilayer graphene metasurface, and
unidirectional propagation of light along the negative direction of the $x$-axis of the domain-wall
interface is observed. Similarly, as shown in Fig.~\ref{fig:fig4}(b), a LCP light source located at
the center of the metasurface, created by reversing the phase difference between adjacent dipoles,
excites at the interface a topological mode that propagates along the positive direction of the
$x$-axis.

\begin{figure}[t!]
    \centering
    \includegraphics[width=\columnwidth]{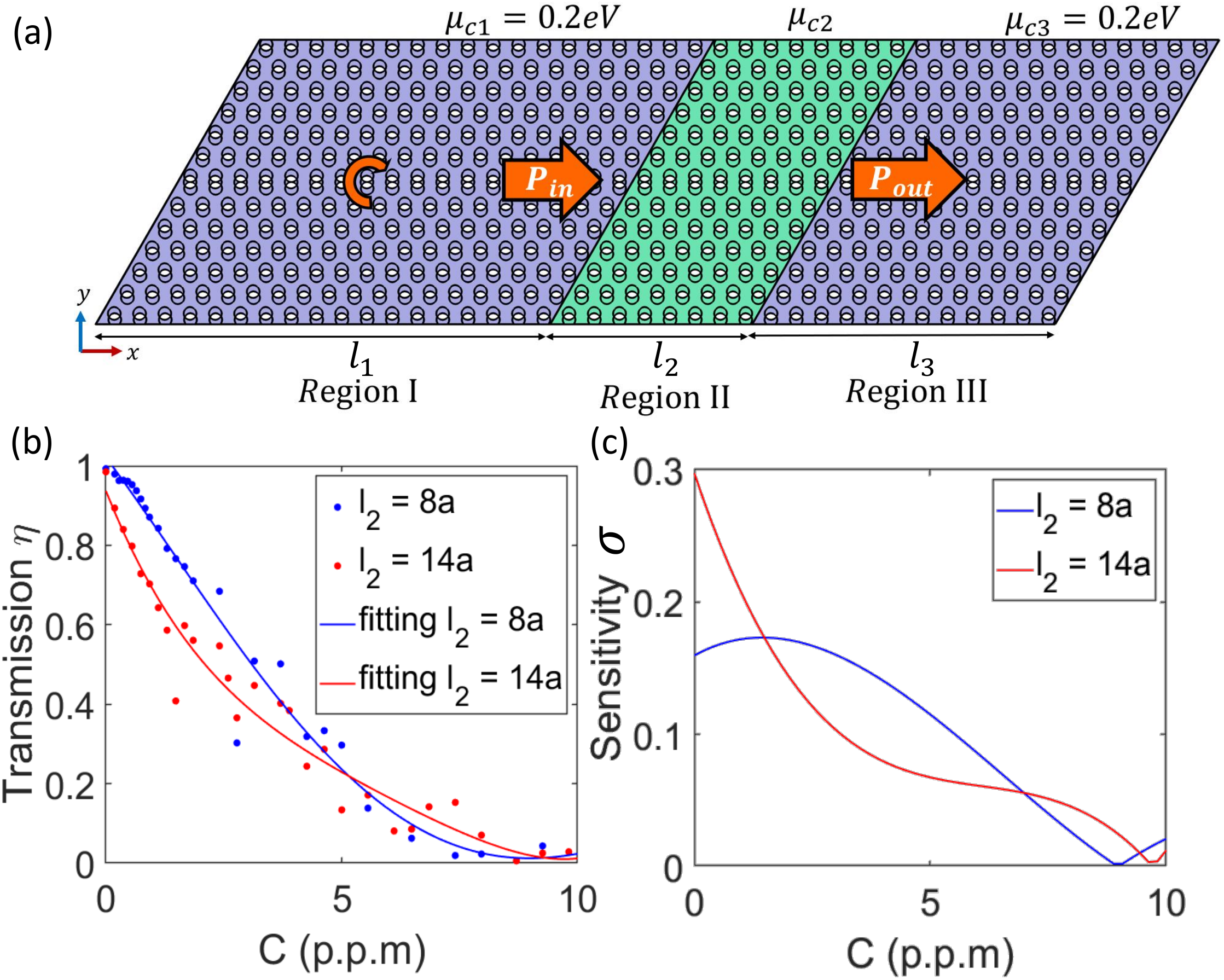}
    \caption{\label{fig:fig5} (a) Schematic of the proposed molecular sensor. The topological
        interfacial mode carries an input power and output power in the regions
        \uppercase\expandafter{\romannumeral1} and \uppercase\expandafter{\romannumeral3}, respectively. An
        additional bilayer graphene metasurface in the region \uppercase\expandafter{\romannumeral2} is
        sandwiched in-between the regions \uppercase\expandafter{\romannumeral1} and
        \uppercase\expandafter{\romannumeral3}, and is used to detect the concentration of adsorbed
        molecules of a certain gas (NO$_2$ in our case). (b) Light transmission, defined as the ratio
        between the output and input power, vs. the concentration $C_{\mathrm{NO_{2}}}$ of NO$_2$ gas,
        determined for $l_2=8a$ and $l_2=14a$. (c) Dependence of the sensitivity of the molecular sensor on
        the concentration $C_{\mathrm{NO_{2}}}$ of NO$_2$.}
    \noindent\rule[0.25\baselineskip]{\linewidth}{0.5pt}
    \vspace{-9mm}
\end{figure}

The unidirectional propagation feature of the topological interface mode of the bilayer graphene
metasurface investigated in this work can find applications to efficient photonic nanodevices. To
illustrate this, in what follows we demonstrate how the interfacial topological mode can be used as
the key component of a molecular sensor. Thus, graphene is a particularly promising 2D material for
sensing applications, chiefly due to its tunable chemical potential and high optical damage
threshold \cite{bao2012graphene}. Generally, the chemical potential $\mu_c$ of graphene is
proportional to the Fermi velocity and the carrier density $n_0$, which can be tuned via molecular
doping, which is a particular type of chemical doping \cite{rodrigo2015mid}.  As such, graphene
based sensors can be used to detect the concentration of specific gases in the environment, by
measuring the concentration of the corresponding molecules adsorbed onto a graphene sheet. To be
more specific, as shown in Fig.~\ref{fig:fig5}(a), the proposed bilayer graphene metasurface is
used to design a molecular sensor based on the large variations of its optical properties induced
by small changes of its chemical characteristics.

The sensor consists of three bilayer graphene metasurfaces, marked as regions
\uppercase\expandafter{\romannumeral1}, \uppercase\expandafter{\romannumeral2}, and
\uppercase\expandafter{\romannumeral3}, the lengths of these regions being $l_1$, $l_2$, and $l_3$,
respectively, and the corresponding chemical potentials $\mu_{c1} = \mu_{c2} = \mu_{c3} =
\SI{0.2}{\electronvolt}$. The gas molecules can be adsorbed only in the region
\uppercase\expandafter{\romannumeral2}, and upon their adsorption $\mu_{c2}$ varies. In practice,
this can be achieved by covering the regions \uppercase\expandafter{\romannumeral1} and
\uppercase\expandafter{\romannumeral3} with some material, e.g., polymethyl methacrylate -- PMMA.
To add specificity to our analysis, we assume that the gas is NO$_2$. The relation between the
variation of the chemical potential induced by NO$_2$ gas with concentration, $C_{\mathrm{NO_2}}$,
is $\Delta\mu_c=\alpha C_{\mathrm{NO_2}}$, where the experimentally determined value of $\alpha$ is
$\alpha\approx\SI{5.4e-3}{\electronvolt}$/p.p.m
\cite{schedin2007detection,hu2019gas,novoselov2005two}. The variation of the chemical potential in
the region \uppercase\expandafter{\romannumeral2}, in turn, leads to a variation of the graphene
permittivity, and consequently to a shift of the frequency of the topological band gap associated
to region \uppercase\expandafter{\romannumeral2}. This means that, if the frequency of the input
light in region \uppercase\expandafter{\romannumeral1} is in the band gap of this region, the
corresponding topological interfacial mode can be switched to a leaky bulk mode in the region
\uppercase\expandafter{\romannumeral2}. Since the leaky bulk modes are particularly lossy, the
output power $P_{out}$ collected in the region \uppercase\expandafter{\romannumeral3} will sharply
decrease.

In order to validate these ideas, we have studied the light transmission in the proposed graphene
metasurface based molecular sensor. To this end, a monochromatic light source with frequency of
\SI{14.16}{\tera\hertz} is placed in the center of the region
\uppercase\expandafter{\romannumeral1}. The lengths of the region
\uppercase\expandafter{\romannumeral1} ($l_1$) and region \uppercase\expandafter{\romannumeral3}
($l_3$) are $18a$ and $12a$, respectively. Moreover, we computed the transmission of the optical
power, $\eta$, defined as the ratio between the output power, $P_{out}$, collected in the region
\uppercase\expandafter{\romannumeral3} and the input power $P_{in}$ in the region
\uppercase\expandafter{\romannumeral1}, namely $\eta=P_{out}/P_{in}$. These calculations were
performed for two different values of the length of region \uppercase\expandafter{\romannumeral2},
namely for $l_2=8a$ and $l_2=14a$, and the corresponding results are summarized in
Fig.~\ref{fig:fig5}.

It can be seen in Fig.~\ref{fig:fig5}(b), where we plot the dependence of the transmission on the
concentration of molecules adsorbed in region \uppercase\expandafter{\romannumeral2}, that the
transmission $\eta$ decreases steeply when the concentration $C_{\mathrm{NO_2}}$ of the NO$_2$ gas
adsorbed in this region increases. In Fig.~\ref{fig:fig5}(b), the dots represent the numerically
computed data, whereas the solid lines indicate the fitting of the results via a third-order
polynomial. These results prove that, as expected, the longer the length of the region
\uppercase\expandafter{\romannumeral2} is, the larger the slope of the transmission curve is, which
means that the radiation loss of the input power in the region
\uppercase\expandafter{\romannumeral2} is larger. Note also that when $C\textgreater\num{5}$
p.p.m., the transmission in the case when $l_{2}=14a$ is larger than when $l_{2}=8a$, which is
attributable to the constructive interference of the mode propagating in region
\uppercase\expandafter{\romannumeral2}, and which undergoes multiple reflections at the interfaces
between this region and regions \uppercase\expandafter{\romannumeral1} and
\uppercase\expandafter{\romannumeral3}. When the concentration $C_{\mathrm{NO_2}}$ is larger than
about \num{9} p.p.m, most of the input power is scattered out into radiation modes, so that the
transmitted power is almost zero in this case. Moreover, we have also studied the sensitivity of
the metasurface sensor, $\sigma$, which is defined as the absolute value of the first-order
derivative of the transmission with respect to the concentration of NO$_2$ molecules adsorbed in
region \uppercase\expandafter{\romannumeral2}, that is, $\sigma=\vert d\eta/dC_{NO_2}\vert$. As
shown in Fig.~\ref{fig:fig5}(c), the proposed molecular sensor can be used to detect the gas
variations in a broad range of molecular concentrations, its sensitivity being particularly large
for small concentrations of adsorbed molecules.

In conclusion, we have proposed a novel mechanism to realize valley-Hall topological plasmon
transport in a bilayer graphene metasurface. In order to create a topological nontrivial valley
bandgap, the lattice of holes of the top layer of the two freestanding graphene layers is
horizontally shifted by a certain distance with respect to the bottom layer, such that the mirror
symmetry between the top and bottom layers is broken. Moreover, to produce a valley-Hall
topological plasmon mode within the nontrivial bandgap, a domain-wall interface is constructed by
placing together two bilayer graphene metasurfaces in a way in which the composite metasurface is
mirror-symmetric with respect to the interface. The results of our numerical computations show that
the proposed domain-wall interfacial waveguide supports topological modes that exhibit
unidirectional propagation feature. This property is further used to design a molecular sensor
based on the fact that the chemical potential of graphene can by tuned \textit{via} gas molecule
adsorption. Our work could have an important impact on the development of integrated plasmonic
devices and key applications pertaining to molecular sensing.

\medskip

\noindent\textbf{Funding.} European Research Council (ERC) (ERC-2014-CoG-648328); China Scholarship Council (CSC); University College London (UCL).

\medskip

\noindent\textbf{Disclosures.} The authors declare no conflicts of interest.


\begin{thebibliography}{99}
    \bibitem{lu2014topological}L. Lu, J. D. Joannopoulos, and M. Soljacic, Nat. Photonics \textbf{8},
    821 (2014).

    \bibitem{khanikaev2017two}A. B. Khanikaev and G. Shvets, Nat. Photonics \textbf{11}, 763 (2017).

    \bibitem{ozawa2019topological}T. Ozawa, H. M. Price, A. Amo, N. Goldman, M. Hafezi, L. Lu, M. C.
    Rechtsman, D. Schuster, J. Simon, O. Zilberberg, and I. Carusotto, Rev. Mod. Phys. \textbf{91},
    015006 (2019).

    \bibitem{niu2007prl}D. Xiao, W. Yao, and Q. Niu, Phys. Rev. Lett. \textbf{99}, 236809 (2007).

    \bibitem{shvets2016njp}T. Ma and G. Shvets, New J. Phys. \textbf{18}, 025012 (2016).

    \bibitem{zhang2017ol}W. Zhang, X. Chen, and F. Ye, Opt. Lett. \textbf{42}, 4063 (2017).

    \bibitem{you2020}J. W. You, Z. Lan, Q. Bao, and N. C. Panoiu, IEEE J. Sel. Top. Quantum Electron. In print, DOI: 10.1109/JSTQE.2020.2982991 (2020).


    \bibitem{jin2017infrared}D. Jin, T. Christensen, M. Soljacic, N. X. Fang, L. Lu, and X. Zhang,
    Phys. Rev. Lett. \textbf{118}, 245301 (2017).

    \bibitem{ylp20sa}J. W. You, Z. Lan, and N. C. Panoiu, Sci. Adv. \textbf{6}, eaaz3910 (2020).

    \bibitem{you2018nonlinear}J. W. You, S. Bongu, Q. Bao, and N. C. Panoiu, Nanophotonics \textbf{8},
    63 (2018).

    \bibitem{gonccalves2016introduction}P. A. D. Goncalves and N. M. Peres, \textit{An Introduction to
        Graphene Plasmonics} (World Scientific, 2016).

    \bibitem{singh2011graphene}V. Singh, D. Joung, L. Zhai, S. Das, S. I. Khondaker, and S. Seal, Prog.
    Mater. Sci. \textbf{56}, 1178 (2011).

    \bibitem{zhang2013review}Y. Zhang, L. Zhang, and C. Zhou, Acc. Chem. Res. \textbf{46}, 2329 (2013).


    \bibitem{kittel1996introduction}C. Kittel, P. McEuen, and P. McEuen, \textit{Introduction to solid
        state physics}, $8^{\textrm{th}}$ ed. (Wiley New York, 1996).


    \bibitem{xie2018photonics}B. Y. Xie, H. F. Wang, X. Y. Zhu, M. H. Lu, Z. Wang, and Y. F. Chen, Opt.
    Express \textbf{26}, 24531 (2018).

    \bibitem{tsofe2007pre}Y. J. Tsofe and B. A. Malomed, Phys. Rev. E. \textbf{75}, 056603 (2007).

    \bibitem{ha2008josa}S. Ha and A. A. Sukhorukov, J. Opt. Soc. Am. B \textbf{25}, C15 (2008).

    \bibitem{bao2012graphene}Q. Bao and K. P. Loh, ACS Nano \textbf{6}, 3677 (2012).

    \bibitem{rodrigo2015mid}D. Rodrigo, O. Limaj, D. Janner, D. Etezadi, F. J. G. De Abajo, V. Pruneri,
    and H. Altug, Science \textbf{349}, 165 (2015).

    \bibitem{schedin2007detection}F. Schedin, A. Geim, S. Morozov, E. Hill, P. Blake, M. Katsnelson,
    and K. Novoselov, Nat. Mater. \textbf{6}, 652 (2007).

    \bibitem{hu2019gas}H. Hu, X. Yang, X. Guo, K. Khaliji, S. R. Biswas, F. J. G. de Abajo, T. Low, Z.
    Sun, and Q. Dai, Nat. Commun. \textbf{10}, 1 (2019).

    \bibitem{novoselov2005two}K. S. Novoselov, A. K. Geim, S. Morozov, D. Jiang, M. I. Katsnelson, I.
    Grigorieva, S. Dubonos, and A. A. Firsov, Nature \textbf{438}, 197 (2005).

    \bibitem{anker2010biosensing}J. N. Anker, W. P. Hall, O. Lyandres, N. C. Shah, J. Zhao, and R. P.
    Van Duyne, in \textit{Nanoscience and Technology: A Collection of Reviews from Nature Journals}
    (World Scientific, 2010) p. 308.

\end{thebibliography}
\end{document}